# Anomalous Hall Effect in Variable Range Hopping Regime: Unusual Scaling Law and Sign Reversal with Temperature

R. M. Qiao[1,2*], S. S. Yan[1*], T. S. Xu[1], M. W. Zhao[1], Y. X. Chen[1], G. L. Liu[1], W. L. Yang[2], R. K. Zheng[3], and L. M. Mei[1]

[1]School of Physics, National Key Laboratory of Crystal Materials, Shandong University, 27 Shanda Nanlu, Jinan, Shandong, 250100, China

[2]Advanced Light Source, Lawrence Berkeley National Laboratory, MS6R2100, One cyclotron road, Berkeley, CA 94720

[3]Australian Centre for Microscopy & Microanalysis, Madsen Building F09, The University of Sydney, NSW 2006, Australia

*To whom correspond:

rqiao@lbl.gov

shishenyan@sdu.edu.cn

**Abstract**

Anomalous Hall effect (AHE) is important for understanding the topological properties of electronic states, and provides insight into the spin-polarized carriers of magnetic materials. AHE has been extensively studied in metallic, but not variable-range-hopping (VRH), regime. Here we report the experiments of both anomalous and ordinary Hall effect (OHE) in Mott and Efros VRH regimes. We found unusual scaling law of the AHE coefficient $R_{AH} \propto \rho_{xx}^{\beta}$ with $\beta > 2$, contrasting the OHE coefficient $R_{OH} \propto \rho_{xx}^{\alpha}$ with $\alpha < 1$. More strikingly, the sign of AHE coefficient changes with temperature with specific electron densities.

Anomalous Hall effect (AHE) typically occurs in magnetic materials with broken time-reversal symmetry as a consequence of spin-orbit coupling[1, 2]. Fundamentally, AHE plays an important role in understanding the topological properties of electronic states, and quantum AHE has recently been found in topological insulators[3]. Technologically, AHE is one of the most critical measurements to probe spin-polarized carriers in magnetic materials for spintronic applications[4]. In particular, the sign and the scaling law of the intrinsic AHE coefficient $R_{AH} \propto \rho_{xx}^{\beta}$ are two key parameters to characterize AHE. Previously, most AHE studies focus on metallic ferromagnets. In the metallic regime, both intrinsic and extrinsic mechanisms, as well as the corresponding scaling law, are well established [1, 5-10]. Comparing with the extensively studied AHE in metallic regime, experimental and theoretical studies of AHE in hopping regime are very limited. Although ferromagnetic semiconductors such as Mn-doped (Ga,Mn)As [4, 11] and transitional-metal-doped oxide [12, 13] have somewhat stimulated the investigation of the AHE in the hopping regime, the weak Hall effect with large longitudinal resistance leads to intrinsic technical difficulty for separating the small Hall signal from the total voltage [14]. Achieving strong AHE in the hopping regime remains a technical challenge, but is critical for studying AHE.

Theoretically, the spin polarization of the weakly localized carriers, the spin-orbit coupling of the localized impurity states at different energy levels, and the specific percolation networks are largely unexplored. Generally, the resistance in hopping regime is dominated by phonon-assisted hopping and has rather distinctive temperature dependence, i.e. the resistivity drops as the temperature rises. The resistivity follows the temperature dependence of $\rho_{xx} \propto \exp[(T_{xx}/T)^{\gamma}]$ with $\gamma = 1/4$ for Mott variable range hopping (VRH) or $\gamma = 1/2$ for Efros VRH[15, 16]. Based on Holstein's work [17, 18], understanding the Hall effect in hopping conduction requires considerations of the interference between direct and indirect hopping via a closed-loop of triads, which gives rise to the microcosmic Hall current [14, 19, 20]. For OHE, the interference is a reflection of the Aharov-Bohm phase due to the external magnetic field, and for AHE, it reflects the Berry phase due to spin-orbit coupling of the spin-polarized electrons. By averaging the Hall current over the conducting network with triads, it was predicted that both OHE and AHE show the same $\exp[(T_0/T)^\gamma]$ dependence as the longitudinal resistivity $\rho_{xx}$. However, the characteristic temperature $T_0$ is smaller than that of longitudinal resistivity [19, 20], leading to the scaling relation of the OHE coefficient $R_{OH} \propto \rho_{xx}^\alpha$ with $\alpha < 1$, and the AHE coefficient $R_{AH} \propto \rho_{xx}^\beta$ with $\beta < 1$.

In this letter, we report both OHE and AHE in the VRH regime realized in a series of $(In_{0.27}Co_{0.73})_2O_{3-v}$ (v denotes the oxygen vacancies) ferromagnetic semiconductors. Electron density in the material is tuned by varying the oxygen vacancy [21, 22]. With certain electron density, the AHE resistivity decreases to zero and then reverses its sign with decreasing temperature. Moreover, an unusual scaling relation of the AHE coefficient displays $\beta > 2$ in contrast with the OHE coefficient $\alpha < 1$, which cannot be explained within existing models. We argue that these surprising results could be related to the localized electrons at different energy levels that are involved in Hall effect in hopping regime.

The $(In_{0.27}Co_{0.73})_2O_{3-v}$ ferromagnetic semiconductor samples are prepared by alternately depositing 0.5nm Co layers and 0.5nm $In_2O_3$ layers for 60 periods on water-cooled glass substrates with magnetron sputtering. In order to tune the electron density, $O_2$ partial pressure is carefully varied from $10^{-5}$ to $10^{-3}$ Pa in the 1 Pa working gas of Ar and $O_2$ [22]. The electron density associated with the oxygen vacancy is measured by ordinary Hall effect at relatively high temperature. By increasing O$_2$ partial pressure for the three samples reported here, A-C, the calculated electron densities are $1.14\times10^{19}$, $6.97\times10^{18}$, and $5.75\times10^{18}$cm$^{-3}$ at 300 K. Combining alternate deposition of layers of atomic thickness and low temperature growth would suppress the crystalline growth and tend to form amorphous compounds[23]. In the amorphous system, the intrinsic structural disorder gives rise to the potential fluctuation as well as the electron localization, which helps maintaining the VRH mechanism in electronic conduction within a wide low temperature range. Therefore, such (In$_{0.27}$Co$_{0.73}$)$_2$O$_{3-v}$ system provides an excellent playground to study the AHE in VRH regime.

The temperature (field) dependence of the longitudinal resistivity $\rho_{xx}(T)$ and the Hall resistivity $\rho_{xy}(H)$ were measured in Van der Pauw configurations by a physical property measurement system (PPMS). All measurements were conducted within linear response regime with current low enough to avoid heating. Since $\rho_{xy}(H)$ is antisymmetric with respect to H but $\rho_{xx}(H)$ is symmetric, the pure $\rho_{xy}(H)$ is able to be obtained by subtracting the contribution of longitudinal resistivity from the raw Hall data. Using the empirical expression $\rho_{xy} = \rho_{OH} + \rho_{AH} = \mu_0 R_{OH} H + \mu_0 R_{AH} M$, the ordinary Hall resistivity $\rho_{OH}$ and the anomalous Hall resistivity $\rho_{AH}$ are well separated from the total Hall resistivity since $\rho_{AH}$ is saturated in high magnetic field. These are standard method for AHE measurement with high accuracy[4]. The magnetic properties were measured by a superconducting quantum interference device (SQUID).

Fig. 1 (a) shows the temperature dependence of the longitudinal resistivity $\rho_{xx}$ measured from 3K to 300K. All samples display semiconducting behavior, namely, $\rho_{xx}$ decreases as

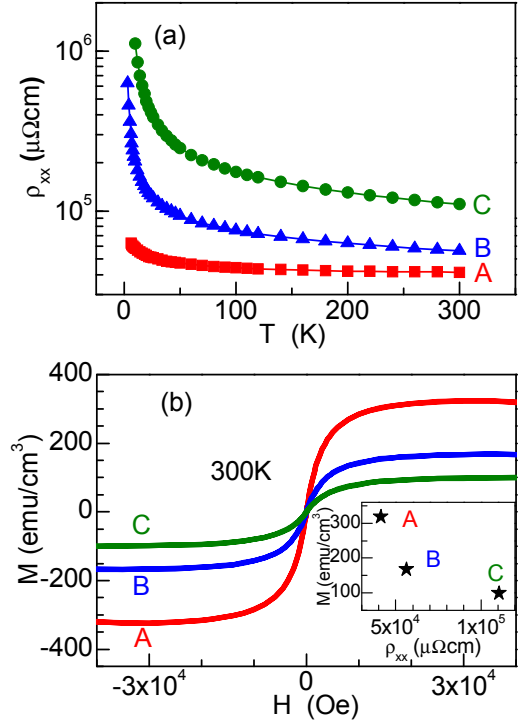

Fig. 1 (color online). (a) Temperature dependence of the longitudinal resistivity $\rho_{xx}$ of the samples A, B and C. (b) Magnetic hysteresis loops of the samples A, B and C measured at 300K. The inset shows the saturated magnetization as a function of the resistivity at 300K for the three samples.

the temperature increases. Fig. 1 (b) shows the magnetic hysteresis loops measured at 300K. The saturated loops and small coercivity indicate that the samples are all ferromagnetic at 300K. The inset in Fig. 1 (b) indicates that the decreasing electron density from A to C leads to more insulating and less ferromagnetic characterizations.

To determine the mode of hopping in our samples, we fit the data to $\rho_{xx}(T) \propto aT^{\lambda} \exp[(T_{xx}/T)^{\gamma}]$, restricting $\lambda = 0$ and $1/n$ with integral values of $n$, and $\gamma$ to 1/4 (Mott VRH[15]), 1/2 (Efros VRH[16]) and 1 (ordinary phonon assisted hopping). We find $T^{\lambda} = 1$ (i.e. $\lambda = 0$) gives the best fitting for all samples[24]. The best description for the temperature dependence of the resistivity is demonstrated in figure 2. The behavior of

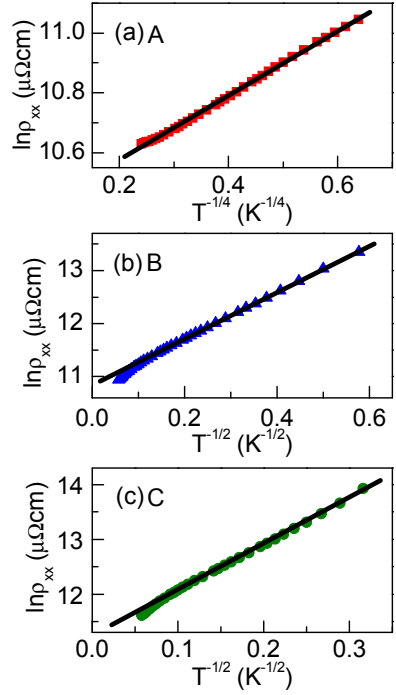

Fig. 2 (color online). $\ln \rho_{xx}$ versus $T^{-\gamma}$. (a) $\gamma = 1/4$ for sample A in Mott VRH regime, (b) $\gamma = 1/2$ for sample B, and (c) $\gamma = 1/2$ for sample C in Efros VRH regime. The straight lines clearly indicate the VRH mechanism at low temperature.

sample A is consistent with Mott VRH, i.e. $\rho_{xx} \propto \exp[(T_{xx}/T)^{1/4}]$. While for sample B and C, Efros VRH, i.e. $\rho_{xx} \propto \exp[(T_{xx}/T)^{1/2}]$ holds at low temperature. The characteristic temperature $T_{xx}$ obtained from the linear fitting of $\ln \rho_{xx}$ versus $T^{-\gamma}$ is listed in table I. The distinct VRH transport in different samples is again due to different electron density associated with oxygen vacancies.

Fig. 3 (a) shows a typical anomalous Hall loop and its corresponding magnetic

Table I. Electrical transport parameters for all three samples. $\gamma$, $T_{xx}$, $T_{OH}$, and $T_{AH}$ have been defined in the text.

| Sample | $\gamma$ | $T_{xx}(K)$ | $T_{OH}(K)$ | $T_{AH}(K)$ | $\alpha = (T_{OH}/T_{xx})^{\gamma}$ | $\beta = (T_{AH}/T_{xx})^{\gamma}$ |
|---|---|---|---|---|---|---|
| A | 1/4 | 1.46 | 0.66 | 755.07 | 0.82 | 4.77 |
| B | 1/2 | 19.40 | 2.95 | 336.59 | 0.39 | 4.17 |
| C | 1/2 | 71.33 | 6.93 | 304.52 | 0.31 | 2.07 |

hysteresis loop measured in sample A at 150K. It is found that $\rho_{AH}$ is proportional to the magnetization as the magnetic field varies. This indicates that the anomalous Hall effect measured by electrical transport method is induced from the same ferromagnetic phase measured by magnetic method like SQUID [25]. The inset in Fig. 3 (a) shows the ordinary Hall resistivity $\rho_{OH}$. The magnitude of $\rho_{AH}$ is around 0.2% of that of $\rho_{OH}$ at the field of 5T. This is in sharp contrast to the Hall effect observed in GaMnAs magnetic semiconductors [11] where AHE usually dominates over OHE. The small magnitude of $\rho_{AH}$ observed in $(In_{0.23}Co_{0.77})_2O_{3-v}$ is typical for wide band-gap oxide magnets [26-28]. It was proposed that the small AHE signal is related with the much smaller spin-orbital coupling in the hopping regime

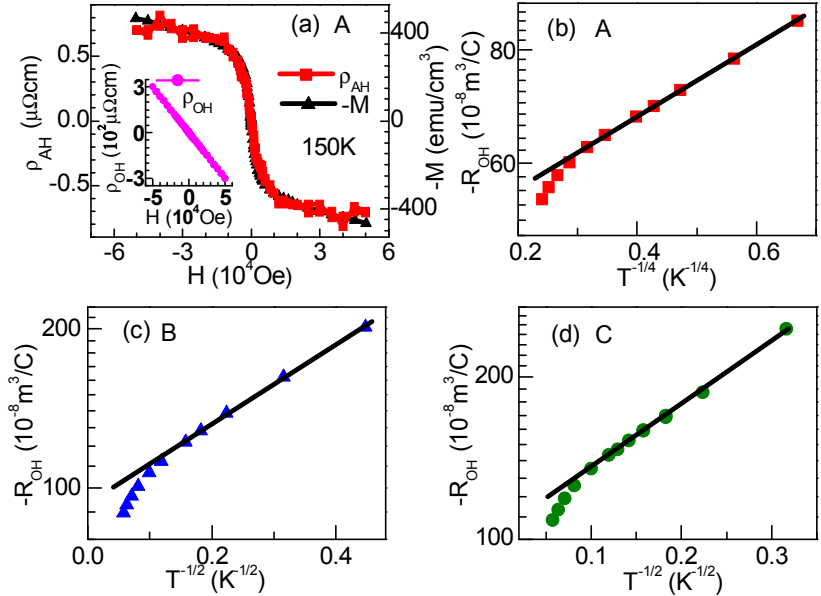

Fig. 3 (color online). (a) A typical anomalous Hall loop and its corresponding magnetic hysteresis loop measured in sample A at 150K. The inset shows the plot of ordinary Hall resistivity $\rho_{OH}$ versus magnetic field H. (b), (c) and (d) show the temperature dependence of the absolute value of the ordinary Hall coefficient, $-R_{OH}$ versus $T^{-\gamma}$, where the vertical scale is logarithmic. (b) $\gamma = 1/4$ for sample A, (c) $\gamma = 1/2$ for sample B, and (d) $\gamma = 1/2$ for sample C, respectively.

of the wide band-gap oxides than that in the extended states[14].

The temperature dependence of the absolute value of the ordinary Hall coefficient, represented by $|R_{OH}| = -\rho_{OH}/\mu_0 H = -R_{OH}$ ($R_{OH}$ is negative for n-type semiconductor) is shown in a logarithmic plot in Figs. 3 (b)-(d). The linear relation between $\ln(-R_{OH})$ and $T^{-\gamma}$ in the low temperature region reveals that the ordinary Hall coefficient also shows VRH behavior like the longitudinal resistivity, i.e, Mott VRH $|R_{OH}| \propto \exp[(T_{OH}/T)^{1/4}]$ for sample A, and Efros VRH $|R_{OH}| \propto \exp[(T_{OH}/T)^{1/2}]$ for samples B and C. Comparing $R_{OH}(T)$ with $\rho_{xx}(T)$, it is easy to find scaling relation of $R_{OH}(T) \propto \rho_{xx}^{\alpha}(T)$ with $\alpha = (T_{OH}/T_{xx})^{\gamma}$, which was listed in table I.

Fig. 4 shows the temperature dependence of the anomalous Hall coefficient $R_{AH}(T)$

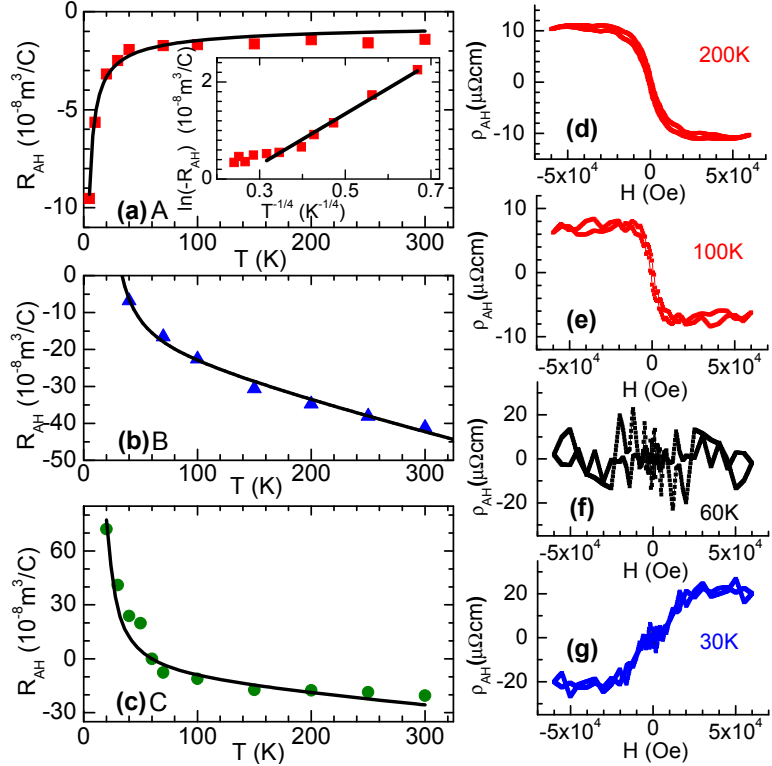

Fig. 4 (color online). Temperature dependence of the anomalous Hall effect. (a)-(c) show the experimental data of $R_{AH}(T)$ versus $T$ and fitting curves. The inset in (a) shows the plot of $\ln(-R_{AH})$ versus $T^{-1/4}$ for sample A, indicating Mott VRH at low temperature. (d)-(g) show the AHE loops of sample C measured at different temperature. The change of hysteresis direction clearly shows the AHE sign reversal.

for samples A, B and C, respectively. At room temperature, the anomalous Hall coefficient $R_{AH}(T)$ is negative in all samples, which is consistent with $R_{OH}(T)$. With decreasing temperature, $R_{AH}(T)$ in sample A (Fig. 4 (a)) monotonously increases in magnitude. In contrast, in sample B, $R_{AH}(T)$ decreases in magnitude with decreasing temperature. Below 40K, the anomalous Hall signal is too weak to be captured by our instrument. In sample C (Fig. 4 (c)), it is surprising that $R_{AH}(T)$ changes its sign from negative to positive with decreasing temperature and the transition temperature is around 60K. The AHE loops measured at different temperatures in sample C are shown in Figs. 4 (d)-(g), which clearly displays the sign change. This is the first report that AHE changes its sign with temperature in the VRH regime. The sign difference of AHE in the VRH regime was observed before only in (Ga, Mn)As samples prepared at slightly different temperatures[11]. According to the hopping AHE theory[14, 19], $R_{AH}$ is proportional to $|d\ln(\rho_0)/d\varepsilon|_{E_F}$, where $\rho_0(\varepsilon)$ is the density of states near Fermi level. Therefore, the AHE is expected to change sign as the Fermi level crosses the density-of-states extremum in the impurity band. Our experimental results show that AHE in the VRH regime is very sensitive to the electronic structure of specific materials. Further theoretical efforts are necessary to clarify the mechanism of our finding of sign reversal with temperature.

We further quantitate the temperature dependence of anomalous Hall coefficient $R_{AH}(T)$. The inset in Fig. 4 (a) shows the plot of $\ln(-R_{AH})$ versus $T^{-1/4}$ for sample A. The linear fitting displays Mott VRH behavior at low temperature, i.e., $R_{AH}(T) \propto \exp[(T_{AH}/T)^{1/4}]$, which yields the Mott characteristic temperature $T_{AH} = 755.07K$ for AHE. Correspondingly, the experimental results and theoretical fitting of $R_{AH}(T)$ in the whole temperature are shown in Fig. 4 (a). Comparing $R_{AH}(T)$ with $\rho_{xx}(T)$, it is easy to find scaling relation of $R_{AH}(T) \propto \rho_{xx}^{\beta}(T)$ with $\beta = (T_{AH}/T_{xx})^{\gamma}$ for sample A, which was listed in table I. For sample C, however, the situation is complicated by the sign reversal. An additional factor $(T-T_{Tr})$, which can be regarded as the first-derivative extension of $|d\ln(\rho_0)/d\varepsilon|_{E_F}$ at transition temperature $T_{Tr}$, is introduced to describe the sign change. The modified Efros VRH behavior is thus given by the expression, $R_{AH}(T) = R_{AH0}(T-T_{Tr})\exp[(T_{AH}/T)^{1/2}]$. The best fitting shown in Fig. 4 (c) (solid line) gives the Efros characteristic temperature $T_{AH} = 304.52K$ and the transition temperature $T_{Tr} = 60K$ for the AHE of sample C. The similar fitting performed on sample B gives $T_{AH} = 336.59K$ and $T_{Tr} = 33.6K$. At temperature far away from $T_{Tr}$, the exponential term $\exp[(T_{AH}/T)^{1/2}]$ dominates the temperature dependence. Therefore the modified scaling relation $R_{AH}/(T-T_{tr}) \propto \rho_{xx}^{\beta}(T)$ is used for sample B and C, where the power index $\beta = (T_{AH}/T_{xx})^{\gamma}$ still holds.

As shown in table I, the values of $\alpha$ and $\beta$ both decrease as the electron density decreases from sample A to C, and it seems to be an universal trend since similar behavior was also found in other systems such as Si:As [29] and GaMnAs [30]. For all samples, it is found that $\alpha < 1$, which suggests that $R_{OH}(T)$ changes less quickly than $\rho_{xx}(T)$ upon temperature. Theoretically, OHE and AHE in the hopping regime originate from the interference between the amplitude for a direct hopping and an indirect hopping via a closed-loop of triads [17, 19]. The observed $\alpha < 1$ in our experiments agrees with the consequence of averaging over the percolation clusters with triads [20, 31], which indicates that models based on such percolation networks are generally valid. However, similar theoretical approach gives anomalous Hall coefficient $R_{AH}(T) \propto \rho_{xx}^{\beta}(T)$ with $\beta < 1$, contrasting the $\beta > 2$ based on experimental data. This unusual scaling law of $\beta > 2$ is a robust result that could be simply obtained by the standard $R_{AH}(T) \propto \rho_{xx}^{\beta}(T)$ fitting for sample A[24]. For sample B and C, the modified scaling relation with $R_{AH}/(T-T_{tr}) \propto \rho_{xx}^{\beta}(T)$ fitting also show $\beta > 2$ [24]. This suggests that $\rho_{AH}(T)$ changes more quickly than $\rho_{xx}(T)$ with temperature in VRH regime. It is worth to mention that the unusual scaling relation with $\beta > 2$ was observed in the metal regime [32], however, the explanation suitable for the metallic regime is not applicable to the hopping regime. The discrepancy between theory and experiments on AHE may stem from the approximation of constant spin-orbit coupling in the existing AHE theory[19, 20]. In the metallic regime, the conduction is due to the conducting electrons near the Fermi energy level. The conducting electron density does not change much with temperature and the spin-orbit coupling energy can be regarded as a constant. In the hopping regime, however, the spin-orbit coupling of the localized electrons at different energy levels is very different. At higher temperature, more localized electrons at deeper energy levels contribute to the hopping. Our experimental results imply that the different spin-orbit coupling energy at different localized energy levels could play an important role and be responsible for the observed unusual scaling relation with $\beta > 2$. In contrast, the spin-orbit coupling has no influence on the OHE, so the OHE coefficient remains $\alpha < 1$.

In summary, we studied both AHE and OHE of a series of $(In_{0.23}Co_{0.77})_2O_{3-v}$ magnetic semiconductors with tunable electron density. These samples enable a systematic experimental study of Hall effect in the VRH regime within a wide temperature range. Both the temperature dependence of OHE coefficient $R_{OH}$ and AHE coefficient $R_{AH}$ show VRH behavior. We discovered an unusual scaling law of the AHE coefficient $R_{AH} \propto \rho_{xx}^{\beta}$ with $\beta > 2$ in contrast with the OHE coefficient $R_{OH} \propto \rho_{xx}^{\alpha}$ with $\alpha < 1$. This is very different from the scaling law in the metal regime. Moreover, with certain electron density, we found the sign of AHE reverses in the VRH regime as temperature varies. We discussed the agreement and discrepancy between our experimental findings and theoretical predictions. The results suggest the general validity of models based on modified Holstein's theory. However the unusual AHE scaling law indicates the different spin-orbit coupling at different localized energy levels may lead to strong temperature-dependence of the AHE coefficient, which has been largely ignored. Our finding of the unusual scaling law, as well as the sign reversal upon temperature, reflects the novelty of AHE in hopping regime, and challenges current understandings of AHE in magnetic materials in semiconducting and insulating regime.

## Acknowledgements:

This work was supported by NSF No.51125004, 111 project B13029, and the National Basic Research Program of China No. 2013CB922303. The Advanced Light Source is supported by the Director, Office of Science, Office of Basic Energy Sciences, of the U.S. Department of Energy under Contract No. DE-AC02-05CH11231.